\documentstyle[11pt]{article}

\newcommand{\ba}{\begin{array}}
\newcommand{\ea}{\end{array}}
\newcommand{\beq}{\begin{equation}}
\newcommand{\eeq}{\end{equation}}
\newcommand{\beqa}{\begin{eqnarray}}
\newcommand{\eeqa}{\end{eqnarray}}

\begin{document}
\begin{center}
{\bf   P-brane solutions in IKKT IIB matrix theory}\\
\vspace{1cm}
Ansar Fayyazuddin \footnote{ansar@nbi.dk}
and
Douglas J. Smith \footnote{D.Smith@nbi.dk}
\\
Niels Bohr Institute, Blegdamsvej 17, DK-2100 Copenhagen {\O}, Denmark
\end{center}

\vspace{.25cm}
\begin{abstract}
We find p-brane solutions to the recently proposed IKKT IIB matrix theory
for all odd p.  We also propose central charges for the p-branes.  
\end{abstract}
\vspace{0.25cm}

NBI-HE-97-02

January, 1997

\vspace{0.50cm}

\section{Introduction}

Recently Banks et.\ al.\ \cite{BFSS} proposed that a non-perturbative
formulation of
M-theory is a certain gauged matrix quantum mechanics.  This theory
is $N=1$ super $SU(N)$ Yang-Mills theory in the large $N$ limit dimensionally 
reduced to one dimension.  
This remarkable proposal has passed a series of tests which suggests
that this may indeed be the correct formulation of the theory.
Inspired by this proposal Ishibashi et.\ al.\ \cite{IKKT} have proposed a matrix
theory (IKKT) in zero dimensions which purports to describe type IIB strings
non-perturbatively.

One of the features of IKKT matrix theory is that it has manifest 10D Lorentz
invariance in contrast to the BFSS description of M-theory which 
does not have manifest 11D Lorentz invariance since it is believed to describe
M-theory in a light cone frame.  
The authors of \cite{IKKT} put their theory to a number of tests.
They showed it supports D-string solutions to its equations 
of motion.  They also  calculated the force between two arbitrarily oriented
static D-strings and showed that it reproduced the result known from 
supergravity.  

\section{p-brane solutions}

As is well known by now, the type IIB string theory has even dimensional 
Ramond-Ramond fields which can couple to odd dimensional $p$-branes. One of
the predictions of string duality is the existence of these $p$-branes.
In this note we show that for every odd integer $p$ there is a $p$-brane
solution to the IKKT matrix model.  We calculate its tension and propose
central charges for each one of the $p$-branes.

We use the slightly modified matrix model proposed by Li \cite{Li}
in the following:
\beq
L = -\frac{1}{2\pi (\alpha ')^{2}g_s}\mbox{Tr }(\frac{1}{4}[ A^{\mu},
A^{\nu}]^{2} + \frac{1}{2}{\bar\psi}\gamma^{\mu}[A_{\mu},\psi ]) 
+\frac{\pi}{g_{s}}\mbox{Tr }1.
\eeq 
Here the $A^{\mu}$ ($\mu =0,...,9$) are hermitian $N\times N$ 
matrices and $\psi$ is
a $N\times N$ matrix of Majorana-Weyl spinors in 10D. We have chosen
a Euclidean formulation for simplicity. To convert to Minkowski space-time
is a simple matter of changing the minus sign in the action to a plus sign
and inserting a factor of $-1$ when raising or lowering the index 0.
The equations of motion with the spinor set to zero are:
\beq
[A_{\mu},[A^{\mu},A^{\nu}] =0.
\eeq
We can construct solutions of this equation such that
\beq
[A^{\mu},A^{\nu}] \propto 1, 
\eeq
and thus commutes with all other matrices.  Such solutions are, as shown
in \cite{IKKT}, BPS saturated (they are annihilated by half of the 
supersymmetry).  
Consider the following solution to the above equations of motion:
\beqa
A_{2k-2} & = &\otimes_{i=1}^{k-1}1_{i}\otimes q_{k}
\otimes_{l=k+1}^{(p+1)/2}1_{l}, \nonumber \\
A_{2k-1} & = &\otimes_{i=1}^{k-1}1_{i}\otimes p_{k}
\otimes_{l=k+1}^{(p+1)/2}1_{l}, \mbox{ for $k=1,..,(p+1)/2$}\nonumber \\
A_{2k-2} & = &A_{2k-1}=0 \mbox{ for $k >(p+1)/2$}.
\eeqa
with 
\beq
[q_{k},p_{k}] =\frac{2\pi i}{n_{k}}T_{k}L_{k}.
\eeq
It should be noted that these operators can only be expressed in terms of
infinite matrices. This can easily be seen since the trace of a commutator
of finite matrices vanishes. However it is convenient to define such
operators formally for finite $n_k$ and then take the limit
$n_k \rightarrow \infty$.
Here the $i$-th sector of the direct product has dimension $n_{i}$.
Hence we have:
\beq
\prod_{i=1}^{(p+1)/2}n_{i} = N,
\eeq
and
\beqa
{[}A_{2k-2},A_{2l-1}] & = & \delta_{kl}
\frac{2\pi i}{n_{k}}T_{k}L_{k}1_{N\times N} \nonumber \\
{[}A_{2k-2},A_{2l-2}] & = & [A_{2k-1},A_{2l-1}] = 0
\eeqa
This solution clearly corresponds to the solution identified as a D-string
in \cite{IKKT} for the particular case of $p=1$.

We substitute the above solution for general $p$ into the action to obtain:
\beq
S_p = \frac{\pi N}{{\alpha '}^2 g_s}\sum_{k=1}^{\frac{p+1}{2}}
	\left ( \frac{T_kL_k}{n_k} \right )^2 + \frac{\pi}{g_s}N
\eeq
We now follow \cite{IKKT} and take the large $N$ limit so that:
\beq
\frac{T_kL_k}{n_k} = c_p \alpha '
\eeq
where we have allowed arbitrary constants $c_p$ for each $p$.
For finite $N$ the system is compactified on a $p+1$ dimensional torus so
we can define the volume as:
\beq
V_{p+1} = \prod_{k=1}^{\frac{p+1}{2}} T_kL_k = (c_p \alpha ')^\frac{p+1}{2} N
\eeq
So we can see that the action reduces to the simple form:
\beq
S_p = \left ( \frac{p+1}{2}c_p^2 + 1 \right )
	\frac{\pi}{g_s(c_p \alpha ')^{\frac{p+1}{2}}} V_{p+1}
\eeq
which we can interpret as the action for a p-brane with tension given by:
\beq
T_p = \left ( \frac{p+1}{2}c_p^2 + 1 \right )
        \frac{\pi}{g_s(c_p \alpha ')^{\frac{p+1}{2}}}
\eeq

Thus we have shown that the correct form of $p$-brane action can be obtained
from the IKKT matrix model for all odd $p$. As we have already noted our
solutions should be D-branes which couple to the Ramond-Ramond fields in
type IIB string theory. This is seen to agree with the formula
for the tension which has the expected $\frac{1}{g_s}$ dependence.
These solutions also preserve half the number of supersymmetries as
D-branes should. The powers of $\alpha '$ are also correctly
produced. However, we cannot predict the exact numerical constants.
We can fix the constants $c_p$ by comparing our result with the tension
calculated from string theory \cite{Polch}. In units where
$\kappa = \gamma \alpha '^2 g_s$ (where $\kappa ^2$ is 8 times the
gravitational constant):
\beq
T_p = \frac{16\pi^{\frac{5}{2}}}{\gamma g_s(4\pi^2 \alpha ')^{\frac{p+1}{2}}}
\eeq

In \cite{IKKT} $N$ was treated as a variable and the equation of motion for
$N$ was used to determine a relation between the unspecified constants in
the original action. Equivalently we can view this equation of motion as
providing the correct scaling relation for $N$ when the action has been fully
determined. A similar procedure here could provide the correct scaling
relations for the $n_k$, determining $c_p$. However, this method does not
seem to be consistent
for $p>1$. One solution to this may be to add higher
order terms to the action. This is not called for if one believes the
deduction of the action from the superstring action. However, if one views
this action as coming from the Eguchi-Kawai reduction of 10D super-Yang-Mills
such terms are to be expected. If suitable terms are introduced with 
undetermined coefficients then consistency of the tensions derived using also
the equations of motion for the $n_k$ would give relations between these
coefficients. In this paper we shall not add higher order terms to the action
and so cannot consider equations of motion for the $n_k$. (However, these
equations are valid and consistent for systems containing only D-strings.)

To show that it is not trivial to identify which $p$-brane(s) some
configuration describes, consider the following:
\beqa
A_0 & = & \left ( \ba{cc} q_1 & 0 \\ 0 & 0 \ea \right ) \nonumber \\
A_1 & = & \left ( \ba{cc} p_1 & 0 \\ 0 & 0 \ea \right ) \nonumber \\
A_2 & = & \left ( \ba{cc} 0 & 0 \\ 0 & q_2 \ea \right ) \nonumber \\
A_3 & = & \left ( \ba{cc} 0 & 0 \\ 0 & p_2 \ea \right ) \label{example}
\eeqa
where
\beq
[q_k,p_k] = \frac{2\pi i}{n_k}T_kL_k1_{n_k \times n_k}
\eeq
with all other commutators zero.

Superficially this looks similar to our solution for a 3-brane (for example
it spans 4 dimensions). However,
there is an important difference since the dimensions of the matrices are
different, i.e.\
$N =$ Tr $1 = n_1 + n_2$
rather than $N = n_1n_2$ for the 3-brane solution.
A little consideration suggests that this solution is a sum of 2 separate
solutions.
If we substitute this solution into the action we will get the sum of 2
D-string actions rather than a 3-brane action. However, for general
configurations it would not always be easy to decide what branes were present.
In the next section we will define central charges which will at least
allow easy determination of the highest dimensional brane present.

\section{Proposed definitions for central charges}

We have shown that there are solutions to the IKKT matrix model which
correspond to D-$p$-branes for all odd $p$. We know from general arguments
that there should be corresponding central charges $Z_p^{\mu \nu \ldots}$ with
$p$ indices. These central charges have the properties that they are
anti-symmetric in all indices and they vanish if any index is $0$
(corresponding to time). Thus we could define totally anti-symmetric objects
with $p+1$ indices $C_{p+1}$ and make the identification:
\beq
Z_p^{\mu \nu \ldots} = C_{p+1}^{0 \mu \nu \ldots}
\eeq

The central charges appear in the supersymmetry algebra but we don't know how
to construct the algebra for the IKKT matrix model. The algebra has been
constructed for the BFSS matrix model \cite{BSS} where it was possible to
define conjugate
momenta but the analogous construction does not seem possible in the
IKKT model. However we will proceed by proposing objects with the correct
properties. The main consideration is that we should define the objects
$C_{p+1}$ so that a $p$-brane has non-zero $Z_p$ and also that $Z_{p'}=0$
for all $p' > p$. In this section we shall always assume $p$ and $p'$ to
be odd integers.

The simplest guess is to define:
\beq
C_{p+1}^{\mu_0 \mu_1 \ldots \mu_p} = A^{[\mu_0}A^{\mu_1} \ldots A^{\mu_p]}
\eeq
where $[\ldots]$ means anti-symmetrise over all indices. We will now show that
this proposed form leads to a consistent definition of central charges, at
least for our D-brane solutions.

First we can show that a $p$-brane has $Z_{p'}=0$ for all $p' > p$. This
follows simply from the definition of $C_{p'+1}$. If for some $\mu$:
\beq
[A^{\mu},A^{\nu}] = 0
\eeq
for all $\nu$
then $C_{p'+1}$ vanishes if it contains the index $\mu$ since $C_{p'+1}$
has an even number of indices. Also $C_{p'+1}$ vanishes if it contains 2
indices the same. But our
solutions obviously contain only $p+1$ fields $A$ which do not commute
with all other $A$'s.
Therefore $C_{p'+1}=0$ and so $Z_{p'}=0$ for all $p' > p$.

Now we can show that $Z_p \ne 0$ for our $p$-brane solutions. First note that
the totally anti-symmetric product of $p+1$ $A^{\mu}$'s can be expressed as a
sum of products of $\frac{1}{2}(p+1)$ commutators $[A^{\mu},A^{\nu}]$ (here
we are using the fact that $p$ is odd for all our solutions). It is now
relatively simple to see that $Z_p \ne 0$ for our solutions. For example if
$p=1$:
\beq
Z_1^{\mu} = [A^0,A^{\mu}].
\eeq
So we have the non-zero component:
\beq
Z_1^1 = 2\pi i c_1 \alpha '
\eeq
Similarly, with a little more effort, we can see that for $p=3$ the non-zero
components of $Z_3$ are given by:
\beq
Z_3^{123} = [A^0,A^1][A^2,A^3] + [A^2,A^3][A^0,A^1] = -8\pi^2(c_3\alpha ')^2.
\eeq
Of course we have not chosen any particular normalisation of $Z_p$ since we
couldn't explicitly construct the central charges from the supersymmetry
algebra. However, with the normalisation chosen above it should now be
clear that our $p$-brane solutions have:
\beq
Z_p^{12 \ldots p} = \left ( \frac{p+1}{2} \right ) !
	(2\pi i c_p \alpha ')^{\frac{p+1}{2}}
\eeq
with the important point not being the actual normalisation but that
$Z_p \ne 0$.

Another observation is that our solutions carry non-zero central charge
$Z_{p'}$ for $p' < p$. This appears to be unavoidable with our construction
but is analogous to the observation \cite{BSS} that a construction of a
longitudinal 5-brane
in the BFSS matrix model carried non-zero $Z_2^{12}$ and $Z_2^{34}$. There
the interpretation was that the longitudinal 5-brane was bound with two
orthogonal infinite stacks of membranes (2-branes). A similar interpretation
in this case would imply that every $p$-brane solution we have constructed is
bound with an infinite number of $p'$-branes for all odd $p' < p$. This is
in a way obvious from the construction of the solutions where we can view any
$p$-brane solution as a $(p-2)$-brane solution at every point of an
orthogonal 2-dimensional space.

One feature which seems to be different in this case compared to the
longitudinal 5-brane case in the BFSS model is that we don't get the
analogous orthogonal stacks of lower dimensional branes. For example our
3-brane solution has only $Z_1^1 \ne 0$. This is due to the
different formulations which treat time very differently.

One should note that a similar phenomenon occurs when there are non-zero
world-volume gauge fields. In this case it was shown in \cite{Douglas} that
$p$-branes carry non-zero charge corrersponding to lower dimensional branes.

It should be noted that our $Z_p$'s are really central charge densities
and are in general matrices rather than c-numbers. In order to calculate
the true central charge we should take the trace of $Z_p$. Apart from
normalisation this is irrelevant in our examples but it is important in
general. For example parallel D-branes can be represented by
block-diagonal matrices and so the trace is necessary to count the
central charge of the system as the sum over each individual D-brane
(block diagonal).

Let us now return to the solution of eqs.~(\ref{example}). Now it is easy
to see that this solution has non-zero $Z_1^1$ as we would expect for a
D-string. Again we see the similarity with the 3-brane solution when we
note that the only non-zero commutators are $[A^0,A^1]$ and $[A^2,A^3]$.
But although these are constants they are not proportional to the identity
in this case. This means that this solution breaks more than half the
supersymmetry. It is also easy to see from the block diagonal structure of
the matrices that $Z_3^{123}=0$ and so this solution
does not have the appropriate charge for a 3-brane.

We have shown how our definitions of central charge can even be used to
identify what certain configurations correspond to. However, there is
still a puzzle about the solution of section~\ref{example}. We have
identified it as a configuration of 2 D-strings yet only one of them
seems to have non-zero $Z_1$. This is explained by the fact that we are working
with a Euclidean metric. This solution really corresponds to 2 completely
orthogonal D-strings in Euclidean space. So when we identify one coordinate
as time and rotate to Minkowski space-time, only one of the D-strings
is physically sensible.

\section{Conclusions}

In this paper we have shown that the IKKT matrix model allows solutions
which can be identified as D-$p$-branes for all odd $p$. We have done this
by showing that the matrix model action reduces to the expected form for
extended objects $S_p = T_p V_{p+1}$. Our solutions also preserve half the
supersymmetry. We have defined central charges $Z_p$ which satisfy the
general properties expected.
The existence of $p$-branes for all odd $p$ can be seen as evidence for
the conjecture that the IKKT matrix model really describes non-perturbative
type IIB string theory.

We have chosen how to take the large $N$ limit by matching tensions to the
expected tensions from string theory. However, it is still not clear why
slightly different limits should be taken for different dimensional objects.

\vspace{0.5cm}
{\bf Acknowledgements}

We would like to thank Y.Makeenko for helpful explanations.

\vspace{0.5cm}
{\bf Note added}

As we were completing this paper the paper \cite{Mak} appeared discussing
similar issues.

\end{document}